\documentclass[letterpaper]{article}
\usepackage{epsf}
\usepackage{theorem}
\newtheorem{definition}{Definition}
\newtheorem{theorem}{Theorem}
{\theorembodyfont{\rmfamily} \newtheorem{qst}{Question}}

\begin{document}
\title{Limits of Rush Hour Logic Complexity}
\author{John Tromp \and Rudi Cilibrasi}

\maketitle

\begin{abstract}
Rush Hour Logic was introduced in \cite{FB99}
as a model of computation inspired by the ``Rush Hour'' toy puzzle,
in which cars can move horizontally or vertically within a parking lot.
The authors show how the model supports polynomial space computation,
using certain car configurations as building blocks
to construct boolean circuits for a cpu and memory.
They consider the use of cars of length
3 crucial to their construction, and conjecture
that cars of size 2 only, which we'll call {\em Size 2 Rush Hour},
do not support polynomial space computation.
We settle this conjecture by showing that the required building blocks
are constructible in Size 2 Rush Hour. Furthermore, we consider
Unit Rush Hour, which was hitherto believed to be trivial, show
its relation to maze puzzles, and provide empirical support for its hardness.
\end{abstract}

\section{Introduction}\label{intro}
Sliding block puzzles are among the simplest kind of motion planning
problems. They involve moving rectangular pieces around inside a bigger
rectangle with the goal of moving a specific piece to a specific target
location. Problems of this kind were first proved to be PSPACE hard in
\cite{HSS84} for arbitrarily large blocks, and later in \cite{HD02}
(or its journal version \cite{HD04}) for fixed size blocks.

A collection of interactive sliding puzzles can be found at
\begin{center}
{\tt http://johnrausch.com/SlidingBlockPuzzles/default.htm}.
\end{center}
This paper focuses on the Rush Hour puzzle, featured at
\begin{center}
{\tt http://www.puzzles.com/products/rushhour.htm},
\end{center}
and quoted as ``one of the most elegant and fun sliding block puzzles
to come on the market in years''.
Its distinguishing characteristic is that the pieces, which are shaped
as cars, can move only in their lengthwise direction, not sideways.
The playing field is a 'parking lot' of size 6 by 6, with
only one exit, and with cars of size 2 by 1 and 3 by 1. The goal is
to get a particular {\em target} car out of the lot through the exit.
In \cite{FB99}, the computational complexity of `Generalized Rush Hour',
played on an arbitrarily large $m$ by $n$ board, is studied. 

\section{Rush Hour Computing}

\begin{figure}
\epsfxsize=10cm \epsfbox{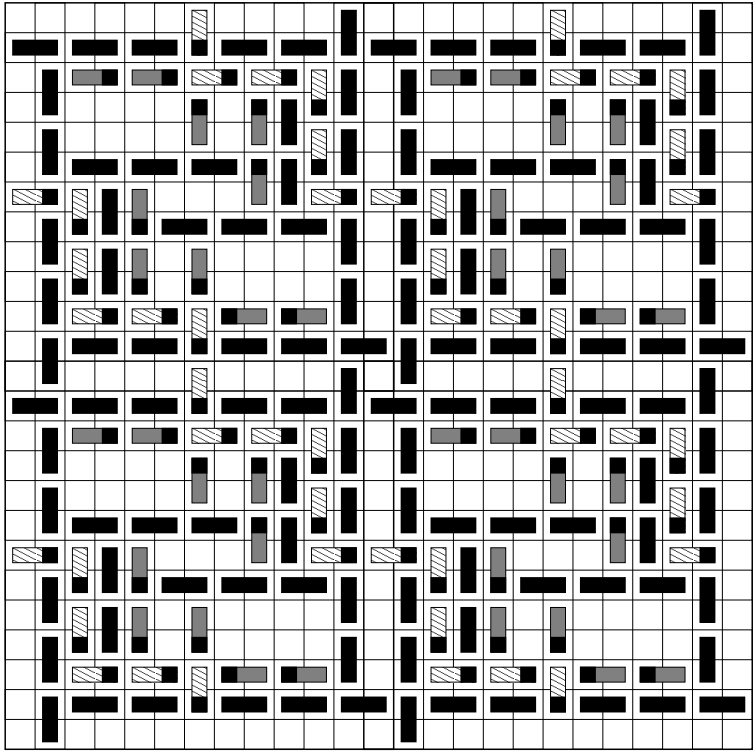} \hspace{1cm}
\caption{four blocks}
\label{blocks}
\end{figure}

Flake and Baum hit upon the brilliant idea of partitioning the whole
board into `blocks' of constant size $k$ by $k$, which are bordered
by walls of interlocking cars, leaving only one gap between adjacent
blocks. Figure~\ref{blocks} shows four such blocks, each of size 11 by 11
(with some cars sticking out at the corners).
Empty space is shown around the cars to make the cars stand out, which
would not be present in the sliding block puzzle. The four black cars
in the center, for example, are completely jammed in and unable to move.
In fact we use black to denote cells that are always occupied by a car.
Cars that can move back and forth one step only, are black at one end,
and gray or striped at the other. The latter is used for chains of
cars connecting the gaps.

Flake and Baum designed particular single block car configurations that
behave as logical gates, imposing certain constraints on the 4 cars
occupying the top, left, right, and bottom gaps of the block.

\subsection{Functional vs Constraint Gates}

Yet these gates are subtly different in behavior from classical
boolean gates, which distinguish one or more inputs from the output.
In classical circuits, the state of a gate's output is uniquely
determined by the state of its inputs,
and wires must always connect an output of one gate to an input of another.

A Rush Hour block instead puts constraints on the combined states of its
`ports'. Each port can be `in' or `out'. The blocks in Figure~\ref{blocks}
forbid the top and right port from both being `in', and also the
left and bottom port from both being `in'.

Consider two adjacent ports, $P$ and $Q$, of two adjacent blocks,
for example the right port of the top-left block and the left port of
the top-right block.

We can consider $P$ an output and $Q$ an input.
The `in' state of $P$ allows $Q$ to be in state `out', which corresponds
to an active wire $P\rightarrow Q$,
since the top-right block now has more freedom
of movement than if $Q$ were `in'.

On the other hand, if we consider $Q$ an output and $P$ an input,
then the `out' state of $Q$ forces $P$ to be in state `in', which corresponds
to an inactive wire $Q \rightarrow P$.

\section{Nondeterministic Constraint Logic}
Hearn and Demaine, in~\cite{HD02}, abstracted the notion of constraining
gates into the `Nondeterministic Constraint Logic' model of computation.
They propose a somewhat abstract graph formulation, as well as a more
concrete circuit formulation for machines, and establish translations
between the two. 
Below I introduce a new formulation which combines the formal rigor of
the graph formulation with the concreteness of the circuit formulation.
I believe the few extra definitions required in this formalism
are offset by the conceptual simplicity and flexibility in gate design.

Basically, a machine is a circuit of gates and wires, where gates
are nodes with labeled {\em half-edges}, and wires
are a matching on half-edges. The half-edges take on the role of ports,
and the `in' or `out' state of a port is naturally represented by
directing the half-edge in or out.

\subsection{Gate types}
\begin{definition}
A gate {\em type} is a tuple $<L,S,E,o>$, where
$L$ is a set of labels,
$S$ a set of states,
$E \subseteq S \times S$ a symmetric set of possible state transitions,
and $o: S \rightarrow 2^L$
gives for each state $s$, an orientation of all half-edges.
Transitions are only allowed between states that differ in the
orientation of exactly one half-edge.
\end{definition}

\begin{figure}
\epsfxsize=10cm \epsfbox{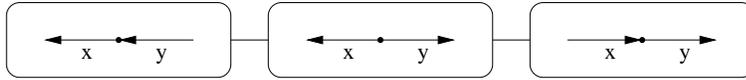} \hspace{1cm}
\caption{a WIRE gate}
\label{wiretype}
\end{figure}

The simplest non-trivial type is the WIRE gate shown in Figure~\ref{wiretype},
which merely forbids its two half-edges to both be inward.

\begin{figure}
\epsfxsize=10cm \epsfbox{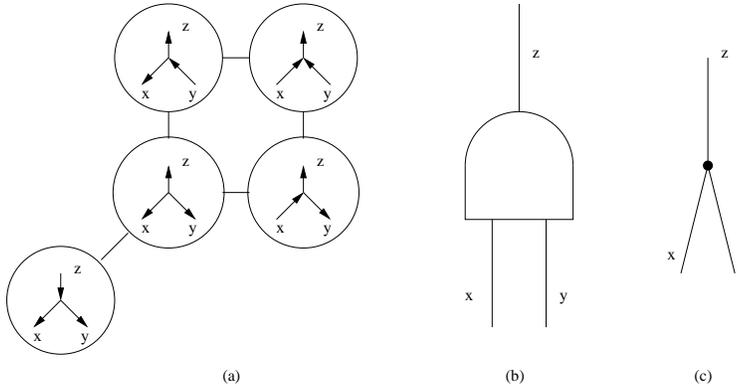} \hspace{1cm}
\caption{(a) gate type AND (b) usual representation (c) viewed as SPLIT}
\label{andtype}
\end{figure}

A more interesting type is the AND gate $<\{x,y,z\},S,E,o>$ whose states,
transitions, and orientations are shown in Figure~\ref{andtype} (a).
The usual representation of an AND gate in Figure~\ref{andtype} (b)
emphasizes the role of $z$ as an output that can be active only when both
inputs $x,y$ are, while the splitting wire emphasizes the role of $x,y$ as
outputs that can be active only when input $z$ is.

\begin{figure}
\epsfxsize=10cm \epsfbox{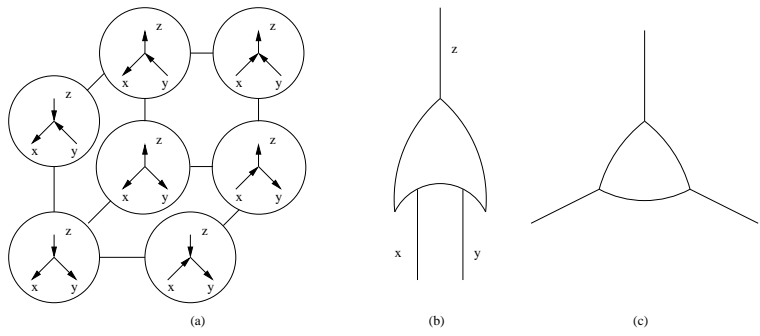} \hspace{1cm}
\caption{(a) gate type OR (b) usual representation (c) symmetric view}
\label{ortype}
\end{figure}

Another type is the OR gate shown in Figure~\ref{ortype} (a).
The usual representation in Figure~\ref{ortype} (b)
arbitrarily shows $x,y$ as inputs and $z$ as output, but from (a)
we can see that $x,y,z$ play equal roles, so a symmetric representation
as in (c) is perhaps more appropriate.

\begin{figure}
\epsfxsize=10cm \epsfbox{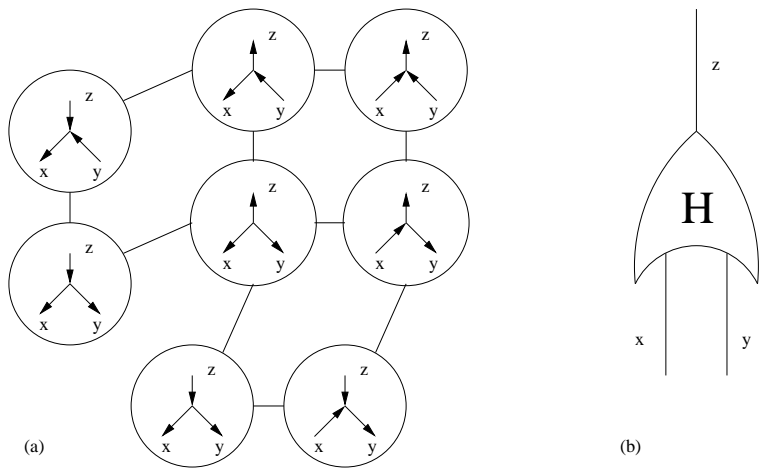} \hspace{1cm}
\caption{(a) gate type HALF-OR (b) representation}
\label{halfortype}
\end{figure}

Our last example is the HALF-OR gate shown in Figure~\ref{halfortype} (a).
This differs from the OR in having two states with both inputs and the output
active. Consider the middle state where both inputs are
active but the output inactive. Then to activate the output, a choice must
be made of which input the output will depend on.
That input then needs to remain active as long is the output is.
For this reason it was called a `Latch' in~\cite{HD02}.
The next section shows the use of the HALF-OR gate.

\subsection{Machines}
A machine is basically a bunch of gates connected together:
\begin{definition}
A machine is an augmented graph in which all half-edges are labeled,
and each node has a gate type consistent with the labels on
its incident half-edges. Half-edges may remain unconnected; these are the
input/output ports of the machine.
\end{definition}

\begin{definition}
A machine state orients all half-edges, consistent across paired half-edges
(one being `in' and the other `out'),
and assigns to each node a state of its gate type consistent with the
incident half-edge orientations.
\end{definition}

\begin{figure}
\epsfxsize=10cm \epsfbox{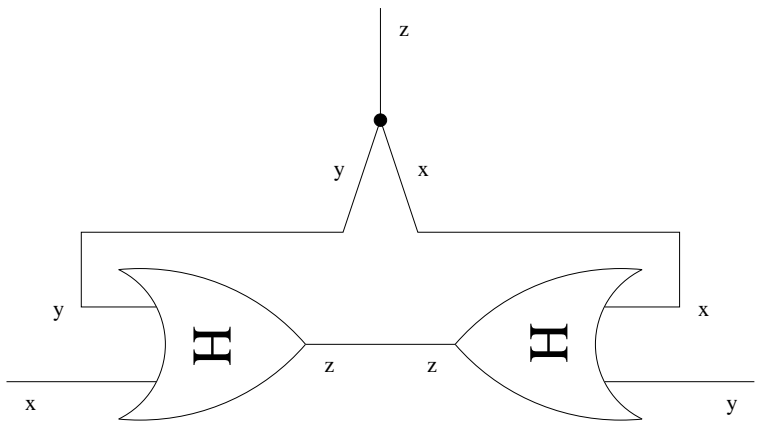} \hspace{1cm}
\caption{machine with one SPLIT and two HALF-OR gates.}
\label{ormachine}
\end{figure}

\begin{figure}
\epsfxsize=10cm \epsfbox{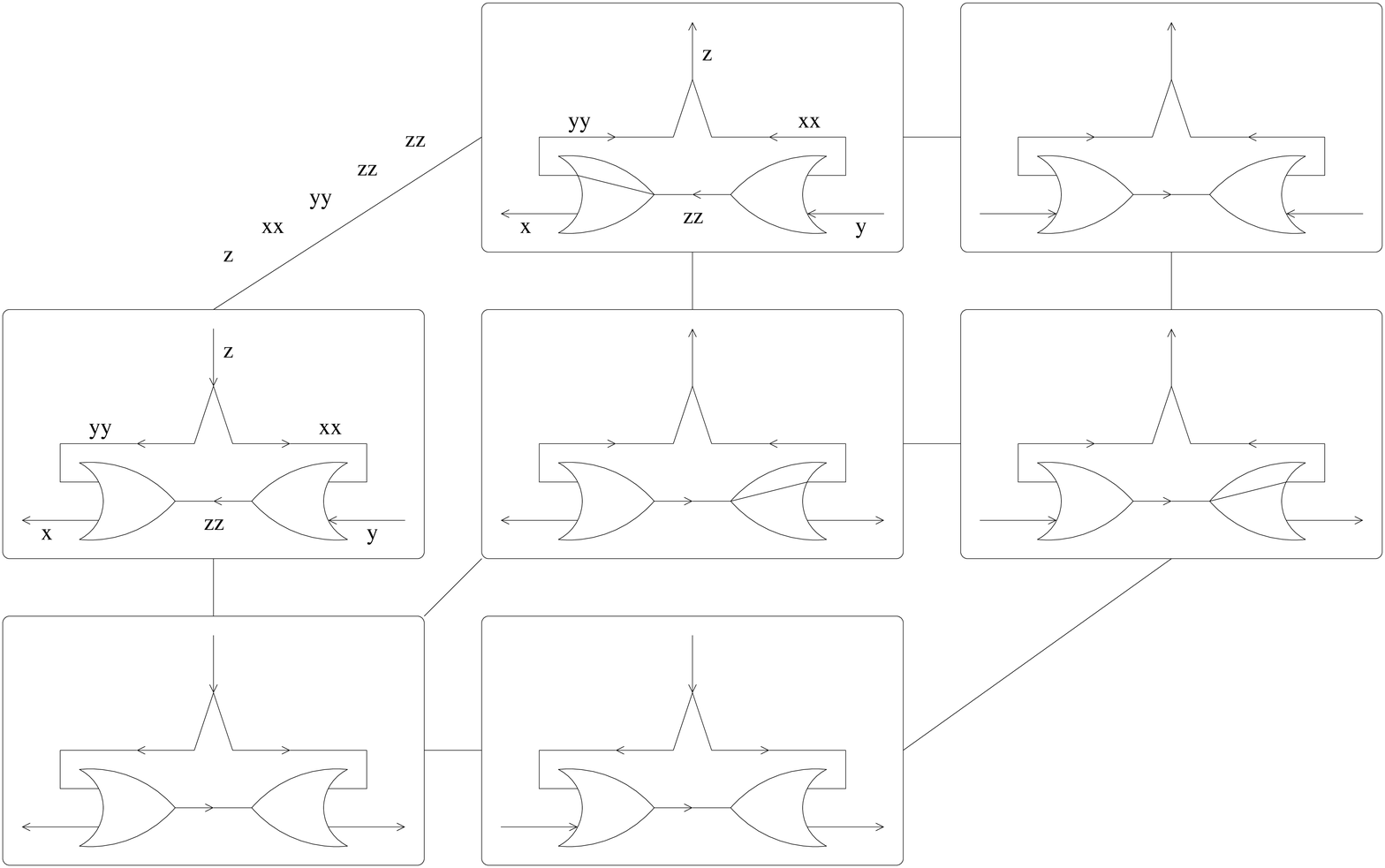} \hspace{1cm}
\caption{possible machine states and multi-step transitions}
\label{orcompose}
\end{figure}

A very simple machine, combining two HALF-ORs with a SPLIT gate, is shown
in Figure~\ref{ormachine}. The fact that we matched same-labeled half-edges
is done just for convenience, so that we may refer to the edges as
$xx$, $yy$, or $zz$.
We may wonder in what state the three unmatched
half-edges can be, among all valid machine states. If $z$ is `in', then
the split requires both the $xx$ and $yy$ edges to be
oriented downward to the half-ors.
Since one HALF-OR must have an in-going $z$, this gate will require an
out-going $x$ or $y$. This argument shows that not all three
{\em machine ports}
can be `in'. If this were the only restriction, then this machine
would seem to behave as an OR gate. Figure~\ref{orcompose} shows that this
is indeed the case. To distinguish the two similarly oriented HALF-OR states,
a line is drawn indicating the input on which the output is dependent.
All 7 states of the OR gate can be achieved as part
of states of the machine, and all 9 transitions are possible as a sequence
of machine transitions. For example, from the middle top state, we can
flip edge $zz$ twice to make the left HALF-OR dependent on $x$ instead of
on $y$, allowing us to flip $yy$, and next flipping $xx$ yields the
machine state at the middle left.

Machines thus allow us to compose new gate types from other ones.
The fact that 2 HALF-ORs and a SPLIT make an OR was first noticed in
an earlier version of this paper, and occurs in~\cite{HD02}
as Lemma 5. Their convention about edge orientation is the opposite
of ours though; a directed edge $s \rightarrow t$ in their paper means
that the wire from output $s$ to input $t$ is active, while in our model
it means that cars have moved out at $a$ and in at $t$, which, considered
as a wire from output $s$ to input $t$, is inactive.

The main result of~\cite{HD02} is that Nondeterministic Constraint Logic,
in particular the question of whether one machine
state can change into another through a series of edge flips,
is PSPACE complete. They even strengthen this to hold for
ternary (degree 3) planar machines, by showing how a CROSSOVER,
essentially a cross product of two WIREs as embedded in a plane,
can be composed of ANDs and ORs.


This general result allows them to give an alternative proof
to the~\cite{FB99} result that Rush Hour is PSPACE complete.
They exhibit 7x7 blocks corresponding to gate types AND and HALF-OR,
and then put 5x5 of these together to make superblocks for
AND, HALF-OR, straight wire, and turning wire, from which arbitrary
ternary planar machines may be built.

\section{Size 2 Rush Hour}
In this section we show our main result that
\begin{theorem}
Size 2 Rush Hour is PSPACE complete.
\end{theorem}

Just as~\cite{HD02} did for Size 2-or-3 Rush Hour, we prove this result
by providing the building blocks to implement any planar ternary
Nondeterministic Constraint Logic machine as an instance
of Size 2 Rush Hour.

\begin{figure}
\epsfxsize=10cm \epsfbox{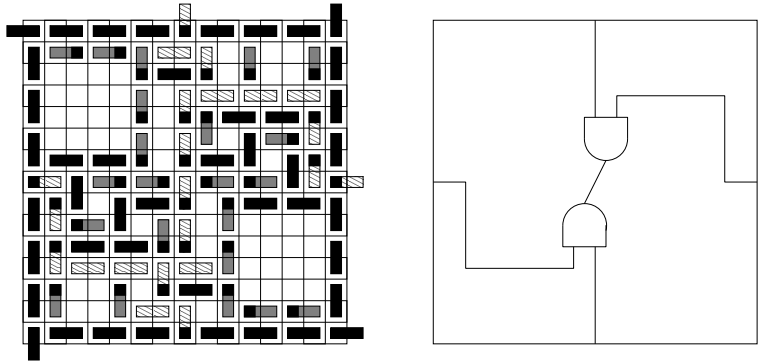}
\caption{AND/SPLIT block}
\label{andblock}
\end{figure}

The block in Figure~\ref{andblock} implements 2 AND gates with their
outputs matched. Omiting the car marked 'A' has the effect of short-circuiting
the top-right AND, leaving one with a single AND and one unconstrained port.

\begin{figure}
\epsfxsize=10cm \epsfbox{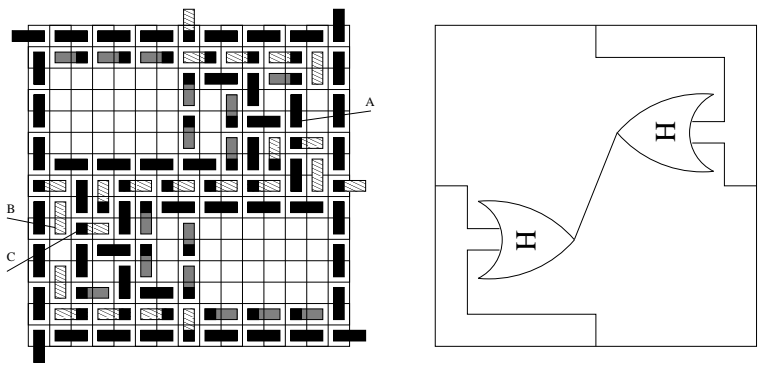} \hspace{1cm}
\caption{HALF-OR block}
\label{orblock}
\end{figure}

The block in Figure~\ref{andblock} implements 2 HALF-OR gates with their
outputs matched. Note that car `C' can be moved
to the left only by locking in car `B', either above or below it.
This corresponds to the dependency we see in the HALF-OR state diagram.

Putting the car marked 'A' in a horizontal position
has the effect of short-circuiting the top-right HALF-OR, while leaving
the top port unconstrained.

\begin{figure}
\epsfxsize=10cm \epsfbox{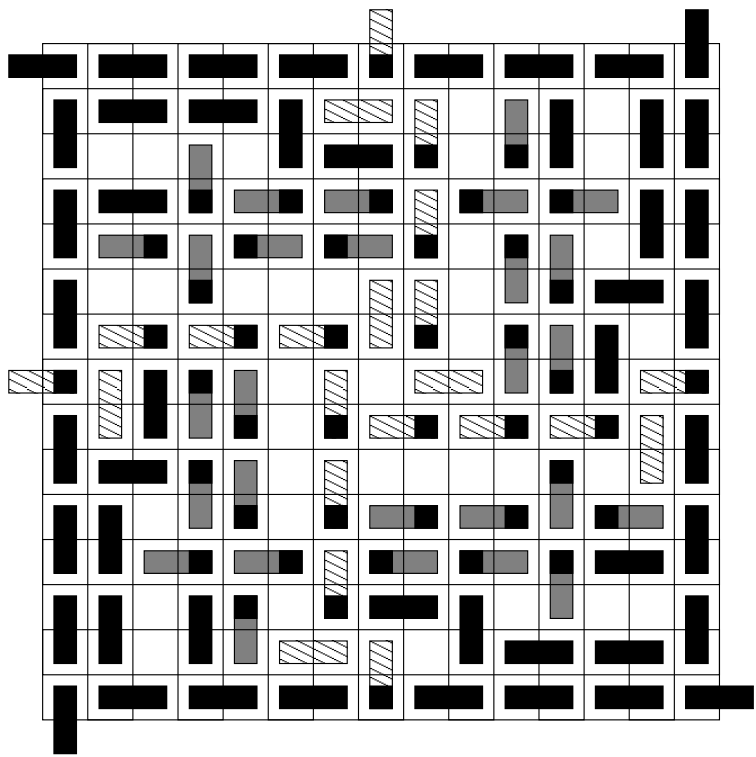} \hspace{1cm}
\caption{CROSSOVER block}
\label{intblock}
\end{figure}

The CROSSOVER block in Figure~\ref{intblock} was designed before the author
learned of its redundancy. Still, it is rather amazing that it can be made
to work in such limited space.
In any case it provides us with a straight WIRE block, which, together
with a stretched out version of the turning WIRE block in Figure~\ref{blocks},
provides us with all the necessary plumbing to connect the logical gates
together.

The proof is completed by verifying the correct operation of each block,
which is greatly facilitated by the coloring. In particular, if it were
possible to vacate any black colored cell, then one such cell would need
to be vacated before any other, and it's easy to manually verify that
this is not possible.

A computer enumeration of all possible states of each block
was used to verify correct operation.

\section{Unit Size Rush Hour}
Taking the constraint on car size to its extreme, we arrive at Unit Size
Rush Hour, where every car occupies exactly one cell.
An example instance is
\begin{verbatim}
||=
|-|
-.|
\end{verbatim}
where `|' denotes a vertical car, `-' and `=' denote horizontal cars, and `.'
an empty cell. The `=' is the unique {\em target car}.
The question is whether some sequence of car moves allows the target
car to reach the left end of its row where it may exit the parking lot.
For our example the answer is yes, as witnessed by the sequence of moves
\begin{verbatim}
 12   11   10    9    8    7    6    5    4    3    2    1    0
||=  ||=  ||=  ||=  |.=  |=.  |=|  |=|  |=|  |=|  |=|  .=|  =.|
|-|  |-|  .-|  -.|  -||  -||  -|.  -||  -||  -.|  .-|  |-|  |-|
-.|  .-|  |-|  |-|  |-|  |-|  |-|  |-.  |.-  ||-  ||-  ||-  ||-
\end{verbatim}

Above each diagram is shown its distance-to-solve.
In this case there is only one empty cell, which necessarily swaps places
with a car on every move. This gives it the feel of a maze problem.
In fact, consider

\begin{definition}
A Rush Hour Maze is a rectangular grid of cells one of which is the
starting location for the player.
The player can move
either horizontally to a cell he last left horizontally,
or vertically to a cell he last left vertically. 
Every cell except the start is oriented to restrict the first arrival.
The exit of the maze is between two specified neighboring cells.
\end{definition}

Then every Unit Rush Hour instance with one empty cell,
in which the target is the leftmost
horizontal car in its row is equivalent to a Rush Hour Maze instance.
The exit is between the two leftmost cells on the target row. Having
the maze player move between these two cells
is equivalent to moving a horizontal
car between the two cells. By assumption, this must be the target car
reaching the exit.

Having the exit of a Rush Hour Maze between cells is important for
ensuring non-triviality,
since the question of whether the player can reach a given exit cell reduces
to the following question:

In the directed graph that connects from each cell to neighbors
of appropriate orientation, does there
exist a path from the player to the exit cell?

This question can be answered with a straightforward depth-first search,
thus all mazes with exit cells are rather trivial.

Similarly, the question of whether a specific car can be moved at all
reduces to the existence of a path from the player to that car\footnote{
The Conclusion of \cite{HD02} mentions this fact and the open problem
of Unit Rush Hour's hardness, based on discussions with this paper's
first author.}.

\section{Right-hand rule}
\begin{figure}
\epsfxsize=4cm \epsfbox{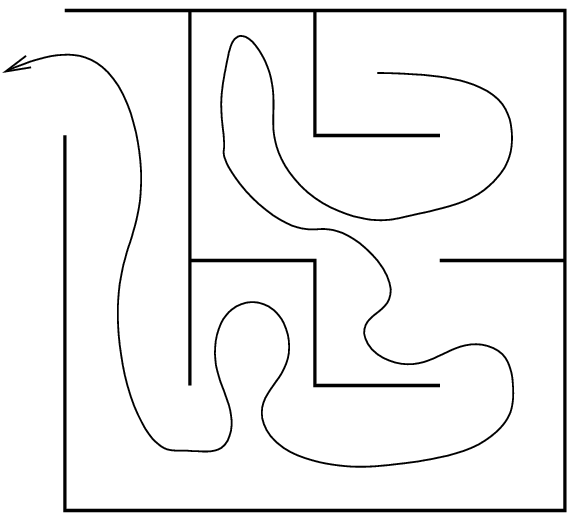} \hspace{1cm}
\caption{solving a 4x4 maze using the right-hand rule}
\label{righthand}
\end{figure}

In plain mazes, one can sometimes follow a simple rule in order to find
the exit. The {\em Right-Hand Rule} says to always
take the rightmost turn, as pictured in Figure~\ref{righthand}.
It works in any maze whose underlying graph is acyclic.
Also in Rush Hour Mazes, the right-hand rule is well defined.
And indeed it is guaranteed to lead to the exit in case the
state graph is acyclic. But whereas the states of a plain maze
are readily apparent as the possible positions of the player,
a Rush Hour Maze can have multiple states with the same player position,
such as positions 3 and 9, or 22 and 32, below.
\begin{verbatim}
  0     3     9    12    14    22    32    36    44
||=-  ||=-  |=-|  |=-|  |=-|  |=-|  ||=-  .|=-  ||=-  
|-|.  |-.|  ||.-  |.|-  |||-  |||-  |-||  |-||  |-|.  
|--|  |-|-  |-|-  ||--  |-.-  --.|  --.|  |--|  |--|  
---|  ---|  ---|  ---|  ---|  |---  |---  |---  ---|  
\end{verbatim}
The right-hand rule fails to find the ``exit'' here, having
pushed the target car back between moves 22--32 leaving it
out of reach in position 36.

\section{Empirical Results}
The most straightforward way to get a feel for the hardness of puzzles
is to study how the worst-case solution length grows with the problem
size. Superpolynomial growth is a prerequisite for hardness.
In order to study solution length in small cases,
we implemented an exhaustive state-space search program for Unit Rush Hour,
focusing on instances with a single empty cell.
Consider the graph $G$ of all possible configurations of width $w$
height $h$ and an exit on row $e$.
With $wh$ possibilities for the location of the empty cell, and $2^{wh-1}$
possible orientations of the cars in the other cells, $G$ has $wh2^{wh-1}$
nodes, or {\em states}.
Searching $G$ requires knowing which states we've visited before.
Storing even one bit per state is undesirable since it would put the case
$w=h=6$ out of reach at 144 Gbyte.
Instead we partition $G$ as follows.
First, we remove from $G$ all states with no horizontal cars on the exit row.
In all remaining states, the leftmost car on the exit row is the target car.
Depending on whether this car in the leftmost cell, a state is either
{\em solved} or {\em unsolved}. Of the solved states, we only keep the
{\em justsolved} ones, where the empty cell is right next to the target car.
This partitions $G$ into many connected components, and preserves the
largest distance to solution.
We can now enumerate all justsolved states, checking if we've visited
it before, and if not, search the whole connected component marking
all encountered justsolved states as visited. This takes only
$2^{wh-2}$ bits, which amounts to a feasible 2 Gbyte for $w=h=6$.
Searching each component in a breadth first manner further helps to conserve
space. The following table lists the
worst-case solution length for each width and height, over all possible
exit rows and states.

\begin{center}
\begin{tabular}{|l|r|r|r|r|r|r|r|r|r|}
\hline
\hspace{5mm} width & 2  &  3  & 4  & 5 & 6 & 7 & 8 & 9 & 10 \\ 
height   &    &     &    &   &   &   &   &   &  \\ \hline
    2  &    3 &  6 &  8 & 10 & 12 & 14 & 16 & 18 & 20 \\ \hline
    3  &    5 & 12 & 21 & 32 & 43 & 54 & 65 & 76 & 87 \\ \hline
    4  &    7 & 21 & 40 & 87 &132 &194 &286 &435 & \\ \hline
    5  &    9 & 31 & 75 &199 &336 &699 & & & \\ \hline
    6  &   11 & 41 &167 &339 &732 & & & & \\ \hline
    7  &   13 & 51 &215 &578 & & & & & \\ \hline
    8  &   15 & 62 &309 & & & & & & \\ \hline
    9  &   17 & 73 &650 & & & & & & \\ \hline
   10  &   19 & 84 & & & & & & & \\ \hline
\end{tabular}
\end{center}

This limited data suggests an exponential growth rate.
It is interesting to analyze these worst-case solutions in detail.

\begin{figure}
\epsfxsize=11cm \epsfbox{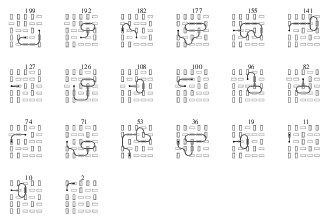} \hspace{1cm}
\caption{solution to hardest 5x5 instance}
\label{sol199}
\end{figure}

Figure~\ref{sol199} visualizes the 199 step solution of the hardest 5x5
instance.
The thick lines indicate the changing position of the empty cell.
Curiously, the solution can be broken down into segments each of which
is either a simple path, or a path followed by a circuit, followed by
the path in reverse. In the latter case, the effect of the segment
is limited to flipping the orientation of all circuit corners,
which somewhat resembles the flipping of a bit.

\begin{figure}
\epsfxsize=11cm \epsfbox{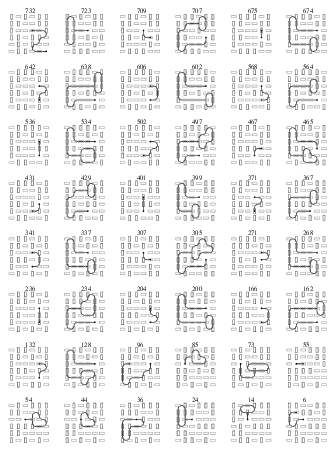} \hspace{1cm}
\caption{solution to hardest 6x6 instance}
\label{sol732}
\end{figure}

Figure~\ref{sol732} visualizes the 732 step solution of the hardest 6x6
instance.

Here, we can also make out `virtual' bit flips on a larger scale. Comparing
the states at distances 674 and 268, we see that they differ in only
the 4 corners of a `virtual' circuit, taking no less than 406 steps to
complete.

There's little hope of exhaustively searching all 7x7 instances
using our approach, as it would take $2^{49-2}$ bits, or 16 Tbyte.
But already in the 6x6 case, a great deal of complexity is apparent.

What is far from apparent though, is how to harness this complexity
in constructing particular circuits, such as binary counters that
would prove the existence of exponential length solutions, or even
the AND and OR gates that provide a basis for
Nondeterministic Constraint Logic. This leaves us with many questions
about the complexity of Unit Rush Hour.

\section{Open Problems}
While Size 2 Rush Hour was shown to be PSPACE complete,
the complexity of Unit Rush Hour eludes us.
It's not clear if limiting the number of empty spaces to one
reduces the complexity of Unit Rush Hour. Empirical results suggest
that the hardest instances do in fact have only one empty space.
Finally, it is possible that Unit Rush Hour becomes more complex if we
can designate some of the cars as being neither horizontal nor vertical
but plain immobile, which is equivalent to saying the parking lot can
have arbitrarily shaped walls. Let's call this generalization
{\em Walled} Unit Rush Hour. Obviously, Rush Hour Maze is no harder
than Unit Rush Hour, which in turn is no harder than Walled Unit Rush Hour.
But this leaves us with a big open

\begin{qst}
What is the complexity of Rush Hour Maze, Unit Rush Hour, and
Walled Unit Rush Hour?
Are they in P, PSPACE complete, or in between?
\end{qst}

\end{document}